# Control of Cu morphology on TaN barrier and combined Ru-TaN barrier/liner substrates for nanoscale interconnects from atomistic kinetic Monte Carlo simulations

*Samuel Aldana\*, Cara-Lena Nies, Michael Nolan\**

Tyndall National Institute, University College Cork, Lee Maltings, Dyke Parade, Cork T12 R5CP, Ireland

E-mail: samuel.delgado@tyndall.ie; michael.nolan@tyndall.ie

ABSTRACT

The miniaturization of electronic devices brings severe challenges in the deposition of metal interconnects in back end of line processing due to a continually decreasing volume available for metal deposition in the interconnect via. Cu is currently used as the interconnect metal and requires barrier and liner layers to prevent diffusion into silicon and promote deposition of smooth conducting metal films, respectively. However, these take up a significant volume in the interconnect vias, which are already extremely narrow due to the high aspect ratio required. Designing combined barrier/liner materials provides opportunities to promote deposition of smooth films and avoid islanding and frees up precious volume to deposit the required amount of interconnect metal. We present here the first *atomistic* kinetic Monte Carlo (kMC) investigation of the potential of Ru-modified TaN as a dual-function barrier/line material to promote 2D morphology of deposited Cu enabling optimization of the limited available interconnect volume. The kMC simulations use Density Functional Theory (DFT) activation barriers to predict the role of the combined barrier/liner material on the morphology of deposited Cu and give, for the first time, a tool to predict metal deposition on technologically important substrates. We study Cu deposition at different partial pressures on a series of substrates, including unmodified TaN and TaN with 25% or 50% surface Ru incorporation, at deposition temperatures of 300 K, 500 K and 800 K. The deposition simulations predict Cu film roughness, island size and number, substrate exposure area (as a measure of 2D *vs.* 3D morphology), layer occupation rate, compactness (as a measure of film quality) and the impact of annealing. As-deposited Cu on unmodified TaN exhibits significantly higher roughness and substrate exposure areas, demonstrating island formation. Ru-modified TaN layers promote horizontal, 2D Cu growth over vertical, 3D island growth. The highest level of Ru incorporation delivers superior film quality and compactness, with substrate exposure reduced by an order of magnitude and reduced roughness compared to unmodified TaN, even at the highest temperatures. Annealing can eliminate vacancy defects and promote horizontal and smoother film growth when the substrate interaction is strong, e.g. 50% Ru-TaN, but it can also promote further island growth on the weakly interacting TaN substrate. Ru incorporation increases the Cu-substrate interaction delivering multiple benefits for Cu deposition by minimizing the number of uncoordinated Cu atoms and reducing the roughness. These results provide valuable insights into the directions needed to optimize metal interconnect deposition on different substrates to achieve improved interconnect quality in electronic applications and showcase a new approach for atomistic simulation of metal deposition on a range of substrates.

INTRODUCTION

The continuous miniaturization of electronic devices, driven by the growing demand for faster but increasingly more energy-efficient information processing technology, has been fueling the semiconductor industry. Device miniaturization enables development of emerging data-intensive applications, such as neuromorphic computing, machine learning algorithm training, the Internet of Things (IoT) and 5G[1]. Although reducing transistor channel lengths to nanometer scales enhances operational speed and transistor density, it also introduces significant performance and reliability challenges. Specifically, the performance of Cu interconnects linking transistors within a circuit does not improve with downscaling, and interconnect manufacture at reduced size becomes increasingly difficult. This poses a severe bottleneck for further miniaturization and critically affects the performance and reliability of nanometer-scale integrated circuits.[2–6]

Notably, the advantages of new technology nodes such as Gate All Around (GAA) and Complementary Field Effect Transistors (CFET) and their superior switching speeds are negated if signal propagation delays in transistor interconnects are not correspondingly reduced. For instance, reducing the metal cross-section by reducing the available volume in a via increases interconnect resistance and, consequently, raises the resistance–capacitance (RC) time constant of the metal–dielectric interconnect, resulting in higher propagation delays.[5,7] This RC delay is a key factor that determines the signal propagation speed across interconnects, affecting the overall speed and performance of integrated circuits[6], as well as heat dissipation and energy consumption—critical features for low-power devices. Decreasing resistance and capacitance in existing downscaled technology involves trade-offs, including increased complexity, higher costs, and potential impacts on device performance and reliability. Consequently, new materials and processes will be required to mitigate RC delay while maintaining the performance and reliability of current interconnects.[4]

Cu has been in use as the interconnect metal in integrated circuits for all technology nodes since 1997, due to its superior resistivity, enabling faster signal transmission, and its higher electromigration resistance, which improves reliability, particularly at smaller scales with higher current densities. However, if Cu comes into contact with silicon, it rapidly diffuses into silicon during thermal stress, degrading the electrical properties.[3,5,8,9] To prevent the diffusion, a barrier material such as TaN is required.[2,6,8,10,11] Furthermore, at nanoscale dimensions, where next generation sub-nm technology nodes such as GAA and CFET will dominate, Cu exhibits high resistivity and tends to form non-conducting 3D islands, particularly during thermal treatment.[8,11] This agglomeration of Cu islands on the surface can lead to reliability and integration issues that prevent Cu from serving as a conducting wire.[8,11] To mitigate the agglomeration of Cu, a liner material, also known as a seed or glue layer, such as Co[12–15] and Ru[11,15–20] can be inserted between the substrate and the adlayer to facilitate the electroplating of a smooth Cu thin film in high-aspect ratio interconnect vias.[10,16,19,21] An appropriate liner material between an interconnect metal and a barrier material requires certain characteristics: (1) it should prevent the agglomeration of the interconnect metal through an interconnect-liner adhesion that is stronger than the interconnect-barrier adhesion; and (2) it should prevent agglomeration of the liner material and the diffusion into the interconnect through a stronger adhesion of the liner to the diffusion barrier than to the interconnect metal.[21]

However, as technology miniaturization continues, the available volume for the barrier/liner/interconnect stack becomes increasingly restricted, making deposition using standard methods such as physical vapor deposition (PVD) more challenging[22]. Depositing two additional layers alongside the Cu interconnect in high aspect ratio vias can result in pinch-off or blockage of the via.[11,18] Despite the challenges Cu interconnects face with miniaturization and the advancements in alternative materials[23,24], Cu remains competitive due to its environmental stability, excellent electrical transport properties[25], and seamless integration with existing technologies[26–28]. To address the issue of limited volume, one promising approach is to design a single material that combines the barrier and liner properties required for Cu deposition. Such a combined material, e.g., a doped or modified metal nitride, can be deposited using atomic layer deposition (ALD)[29], which excels at depositing conformal and thin films on a variety of substrates. Incorporation of two metals can be achieved with ALD through laminate doping[30,31]. ALD helps reduce the number of processing steps and the thickness of the barrier/liner layer[17,20,32], thus ensuring that the interconnect via contains as much Cu as possible to maintain low resistivity and extend Cu usage beyond current technology nodes. Typical combined potential barrier/liner materials are composed of a barrier material that is modified by incorporation of another metal species that promotes Cu wetting, such as TaN(Ru)[11,17,18,20], Co(W)[33], 2D $TaS_x$[34] and amorphous $CoTi_x$[35]. This incorporation approach can modulate the substrate-interconnect interaction strength, allowing control over the film growth mode and thus introducing liner properties to the barrier material. This is critical since weak substrate interactions typical of Cu on barrier layers promote 3D island growth, while strong interactions promote 2D layer-by-layer growth[36–39] so that controlling the metal-substrate interaction strength can be used to control the morphology of deposited metals.

First principles Density Functional Theory (DFT) calculations have been used to investigate the adhesion between Cu, TaN and various liner materials[21], the activation energy for migration of Cu on doped TaN[11,17,18], and the behavior of Cu atoms on Cu[40,41] and Ag surfaces[42]. However, due to high computational cost, DFT calculations are limited to short simulation times and small systems (hundreds of atoms) which makes simulation of metal deposition impossible. Conversely, mean-field approximations offer a cost-effective alternative for studying larger systems (thousands of atoms) over longer timescales with fewer resources, at the expense of overlooking the microscopic details[43–45]. An intermediate approach is the kinetic Monte Carlo algorithm, which captures stochastic, microscopic behavior over macroscopic timescales. This technique has proven valuable for investigating the formation and dynamics of 3D islands on weakly-interacting substrates[36,39,46], the morphology of 2D islands[47,48] or structural changes in emerging devices such as memristors[49–52].

We employ in this paper a new kMC algorithm based on DFT results to study Cu deposition on a TaN barrier modified with varying levels of Ru surface incorporation to yield a combined barrier/liner layer and analyze the performance of this new material in controlling Cu morphology. The fundamental chemistry in Cu morphology on modified TaN has been previously studied using DFT calculations[11,17,18], which can elucidate the early stages of Cu film growth and predict, albeit in a very simple model, how the growth mode can be determined by Cu-substrate interactions. However, such models are limited in scale and time and cannot be used to simulate the complex dynamics of a deposition process or the properties of a deposited metal film. By incorporating key atomic level information from DFT calculations into the kMC simulations, we can predict the quality of the Cu film deposited on TaN with different Ru incorporation levels, under back-end-of-line (BEOL) relevant deposition temperatures and partial pressures. We also evaluate key quantities of the deposited film including film thickness, covered substrate area and occupation rate per layer (assessing vertical or layer-by-layer growth and substrate coverage efficiency), Root Mean Square (RMS) roughness (reflecting surface roughness), and the coordination number of Cu atoms (as a measure of defect density), which in combination allow detailed assessment of the morphology. An important advance compared to the literature is that we examine directly the effects of vacuum annealing on the deposited Cu films to explore how annealing can help promote the desired morphology of a metal film. Overall, this work delivers a new atomistic approach to simulate and predict the deposition and morphology of technologically relevant metal films, which will be critical for development of next generation metal interconnects for GAA and CFET devices.

METHODOLOGY

The interaction between deposited Cu and various substrates was previously studied using periodic spin-polarized DFT with the Vienna Ab initio Simulation Package (VASP, version 5.4[53]).[11,17,18] The exchange-correlation functional was approximated using the Perdew–Burke–Ernzerhof (PBE) method.[54] Valence electrons were explicitly expanded in a periodic plane wave basis set with a kinetic energy cut-off of 400 eV, while core-valence electron interactions were treated using projector augmented wave (PAW) potentials.[55] The valence electron configurations used for Ta, N, Ru and Cu were Ta: $6s^25d^3$; N: $2s^22p^3$; Ru: $5s^14d^7$; Cu: $4s^13d^{10}$.

The activation energies for Cu migration on TaN and TaN(Ru) were computed using the climbing image nudged elastic band (CI-NEB) method with 5 images including the starting and ending geometries.[56,57] The activation energies for Cu on crystalline Cu have been determined using molecular dynamics (MD) calculations.[40,41] Further details on the activation energies can be found in Supporting Information, Section 1. The forces acting on unconstrained atoms during geometry relaxation calculations, as well as the NEB forces in the activation energy computations, were converged to 0.02 eV/Å. Full details about the DFT calculations can be obtained in ref. 11.

The combined liner/barrier materials investigated are based on TaN with varying Ru concentrations in the surface layer: bare TaN, 25% Ru (Ru25) and 50% Ru (Ru50).[11] The modification involves substituting Ta sites with Ru in the top layer of TaN. This composition can be achieved through atomic layer deposition of TaN and then introducing a Ru precursor for the final cycles to incorporate Ru into the TaN surface layer. Notably, the distribution of Ru affects the electronic properties and thermal stability of the Cu structure, as the smaller ionic radius of Ru compared to Ta creates surface recesses than can trap Cu atoms, but that also decreases the thermal stability at high Ru concentrations[11,17,18]. Hence, we limited the kMC simulations to Ru concentrations below 50%, which strengthen the Cu-substrate interactions and therefore can promote 2D Cu morphology.[11]

The insights obtained from these DFT calculations, along with previous ab initio studies on homoepitaxial Cu deposition,[40,41] are used in kMC simulations to investigate the deposition of Cu on the aforementioned substrates, assessing their performance as a barrier/liner material. A discrete lattice kMC model was employed to simulate atom-by-atom adsorption and adatom diffusion between neighboring sites. The Cu lattice used is a ($10 \times 10 \times 2$ nm) face-centered cubic (fcc) structure with (111) orientation, Cu lattice constants ($a = b = c = 0.358$ nm) and periodic boundary conditions in the plane. Each Cu atom has twelve nearest neighbors, as shown in Figure 1a. Specifically, six neighbors are in the same plane (blue particles), three are in the upper layer (orange particles) and three are in the lower layer (green particles). This lattice structure exhibits two types of facets: (001) and (111). Figure 1b, 1c, and 1d present a representative simulation of a Cu film deposited on TaN at 300 K after more than 2.3 million steps giving a film thickness of 1 nm. The (111) facet is shown in the x-y top view (Figure 1b) and the y-z cross section (Figure 1c). The (001) facet is depicted in the cross-section of the lattice, rotated 30 degrees around the z-axis (Figure 1d). The observed irregularities and gaps correspond to typical defects encountered during deposition. The main processes in the simulations are the deposition and migration of Cu atoms either on the substrate or on the growing Cu film. We implemented the following conditions for migration and adsorption processes: For migration, Cu atoms must be supported by either the substrate or at least two nearest neighbors, while for adsorption, they must have the substrate or three nearest neighbors. These conditions, free of geometric constraints, allow for the formation of various three-dimensional shapes.

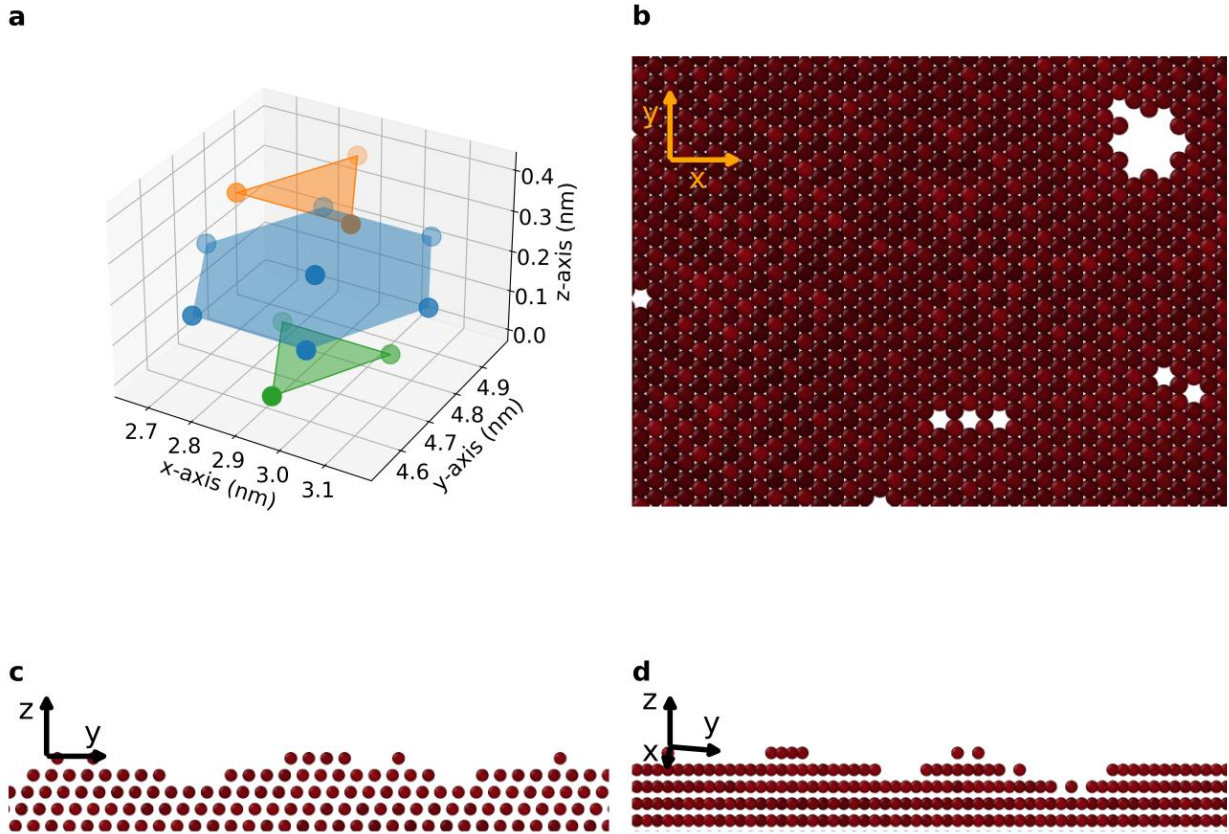

**Figure 1.** a) Atom with its twelve nearest neighbors: six neighbors are in the same plane (blue particles), three are in the upper layer (orange particles) and three are in the lower layer (green particles). A representative simulation of a Cu film deposited on TaN substrate at 300 K after more than 2.3 million steps at partial pressure of 0.1 Pa from different perspectives: b) Top view aligned with z direction (x-y plane) displaying a (111) orientation. c) Cross-section along the x-direction (y-z plane) displaying a (111) orientation. d) Cross-section of the lattice in (001) orientation, following a 30-degree rotation around the z-axis.

To simulate the transitions between different sites in the lattice structure and the relative likelihood of each process, we employ a rejection-free kinetic Monte Carlo algorithm. This methodology involves two steps: 1) computing the transition rates associated with each possible event, and 2) using a random number to select one of these events. The transition rate of Cu migration, a thermally activated process, is determined via transition state theory, dependent on temperature and the activation energy, i.e., $\Gamma = \nu \cdot \exp(-E_A/K_B T)$,[58,59] where $\nu = 7 \times 10^{12}$ s$^{-1}$ is the pre-exponential factor and $E_A$ the activation energy of the corresponding event. In contrast, the transition rate for the deposition process, a non-activated process, is calculated using the kinetic gas theory:[60,61] $k_{ads} = P\sigma(T,\theta)A/\sqrt{2\pi m k_B T}$, where $P$ is the partial pressure of the gas, $\sigma$ is the sticking coefficient that depends on temperature $T$ and surface coverage $\theta$, $m = 63.546$ amu is the molecular mass of Cu, $A$ is the active area and $k_B$ is the Boltzmann constant. The active area A is approximated by dividing the total area of the simulation domain by the number of sites. The sticking coefficient $\sigma$ is set to 1, independent of $T$ and $\theta$, which is a commonly employed approximation[60,62].

The activation energies for Cu migration on various substrates, including TaN and Ru-modified TaN, have been determined using DFT calculations, with migration on TaN surfaces involving a higher activation energy compared to Ru-modified ones.[17] The migration energy barrier for the migration of Cu on TaN substrate ranges from 0.85 to 1.26 eV[17]. In contrast, the barrier is significantly lower on bare hexagonal Ru(001), ranging from 0.07 to 0.11 eV[17]. For simplicity, we assume that the incorporation of Ru into TaN reduces the activation energy for migration on the substrate. Accordingly, we selected three values: 0.85 eV for the bare TaN, 0.6 eV for Ru25 and 0.4 eV for Ru50, which account for differences in Cu migration between the three surfaces found with DFT. As stated previously, the energy barriers for atomic Cu diffusion on a fcc metalic Cu on a (111/001) surface were previously obtained using MD[40,41]. A selection of activation energies of Cu migration processes is provided in Table 1, while further details can be found in Supporting Information, section 1. The activation energies for step ascent, migration from layer n to layer n+1, on various substrates have been selected to be lower than those for homoepitaxial growth of Cu (see Table 1). This is due to the weaker interaction of TaN with Cu, compared to Cu on Cu. Meanwhile, activation energies for Cu to migrate off the substrate and form Cu islands increase with increased Ru incorporation, due to the increased interaction between Cu and Ru-TaN.[11] A bond-counting scheme is implemented to account for changes in the activation energy according to the coordination number (CN) as an atom moves from the initial to the final site. The contribution of the CN to the activation energy is 0.15 eV/atom. Typically, the migration of Cu atoms on a Cu (111) surface maintains the same CN number with 3 neighboring Cu atoms. However, when Cu migrates upward to the next layer (step ascent on Cu film deposited on TaN substrate), their CN changes from 5 Cu neighbor atoms (three in the lower layer and two in the same layer) to only 2 Cu atoms supporting the migration. This transition then results in an activation energy of 0.58 eV, due to the CN change.

The DFT simulations for these substrates showed no migration or exchange of substrate and Cu species after depositing Cu so no migration of N or Ru out of the substrate or Cu into the substrate is allowed in the kMC simulation.

We employ a binary tree data structure to sort the transition rates, facilitating the use of a binary search method for efficiently identifying a chosen event.[63,64] The time step is calculated with a second random number and is weighted by the transition rates: $t = -ln(rand)/\sum \Gamma$, where rand is a random number between 0 and 1 and $\sum \Gamma$ is the summation of the possible events.

**Table 1:** Activation energies for selected Cu migration processes on Cu, TaN, Ru25 and Ru50 substrates.

|  | Cu | TaN | Ru25 | Ru50 |
|---|---|---|---|---|
| Cu Terrace (111) | 0.043[40] | 0.85 | 0.6 | 0.4 |
| Cu Terrace (001) | 0.477[40] | - | - | - |
| Step ascent from substrate (111) | N/A | 0.13 | 0.18 | 0.28 |
| Step descent to substrate (111) | N/A | 0.13 | 0.13 | 0.13 |
| Step ascent from substrate (001) | N/A | 0.19 | 0.23 | 0.38 |
| Step descent to substrate (001) | N/A | 0.318[41] | 0.318[41] | 0.318[41] |
| Step ascent from Cu (111) | 0.313[41] | 0.13 | 0.20 | 0.28 |
| Step descent to Cu (111) | 0.095[41] | 0.095[41] | 0.095[41] | 0.095[41] |
| Along edge Cu (111) | 0.309[41] | - | - | - |
| Along edge Cu (001) | 0.245[41] | - | - | - |

RESULTS AND DISCUSSION

To investigate the impact of different substrates and deposition parameters, we modelled Cu deposition on TaN, Ru25 and Ru50 using the activation energies shown in Table 1 at a series of temperature and partial pressures, keeping within the constraints of BEOL processing.

To evaluate the growth morphology and film quality we study the evolution of key metrics with time and temperature. Specifically, we examined film thickness, extent of covered substrate area and occupation rate per layer to assess competition between vertical and layer-by-layer growth and substrate coverage efficiency. We quantify the extent of island formation and morphology through Root Mean Square (RMS) roughness, island count and the total island mass, and measured defect density by analyzing the coordination number of Cu atoms. In addition, and in contrast to typical kMC simulations of deposition processes, we explored the effects of vacuum annealing on the film properties and their influence on film morphology and quality. Higher quality films are characterized by lower substrate exposure, fewer defects and reduced RMS roughness.

*SURFACE ROUGHNESS AS A MEASURE OF ISLAND FORMATION*

Figure 2 presents the temporal evolution of key film metrics—the mean thickness, the Root Mean Square (RMS) roughness, number of islands and the total island mass—across different substrates (TaN, Ru25 and Ru50), and temperature combinations (300 K, 500 K and 800 K) at a partial pressure of 0.5 Pa during Cu deposition simulations. Simulations were halted when the mean thickness of the deposited film reached 1nm; above this thickness, distinctions between 2D and 3D morphology start to disappear. The resulting deposited films are depicted in Figure S7, with corresponding height color maps in Figure S8. The RMS roughness provides quantitative data regarding the statistical height variations relative to the mean thickness. Islands were identified by locating the highest points in the simulation containing a Cu atom and defining clusters extending down to the mean thickness. The total island mass was calculated by summing the number of atoms within each detected island. For the analysis, only islands containing more than 10 atoms were considered. The study was extended to six different partial pressures (0.1, 0.5, 1, 10, 40 and 100 Pa), revealing that variations in pressure primarily reduce the time scale of the deposition process, while other metrics remained comparable, as illustrated in Figure S9 (0.1, 0.5 and 1 Pa) and S10 (10, 40 and 100 Pa).

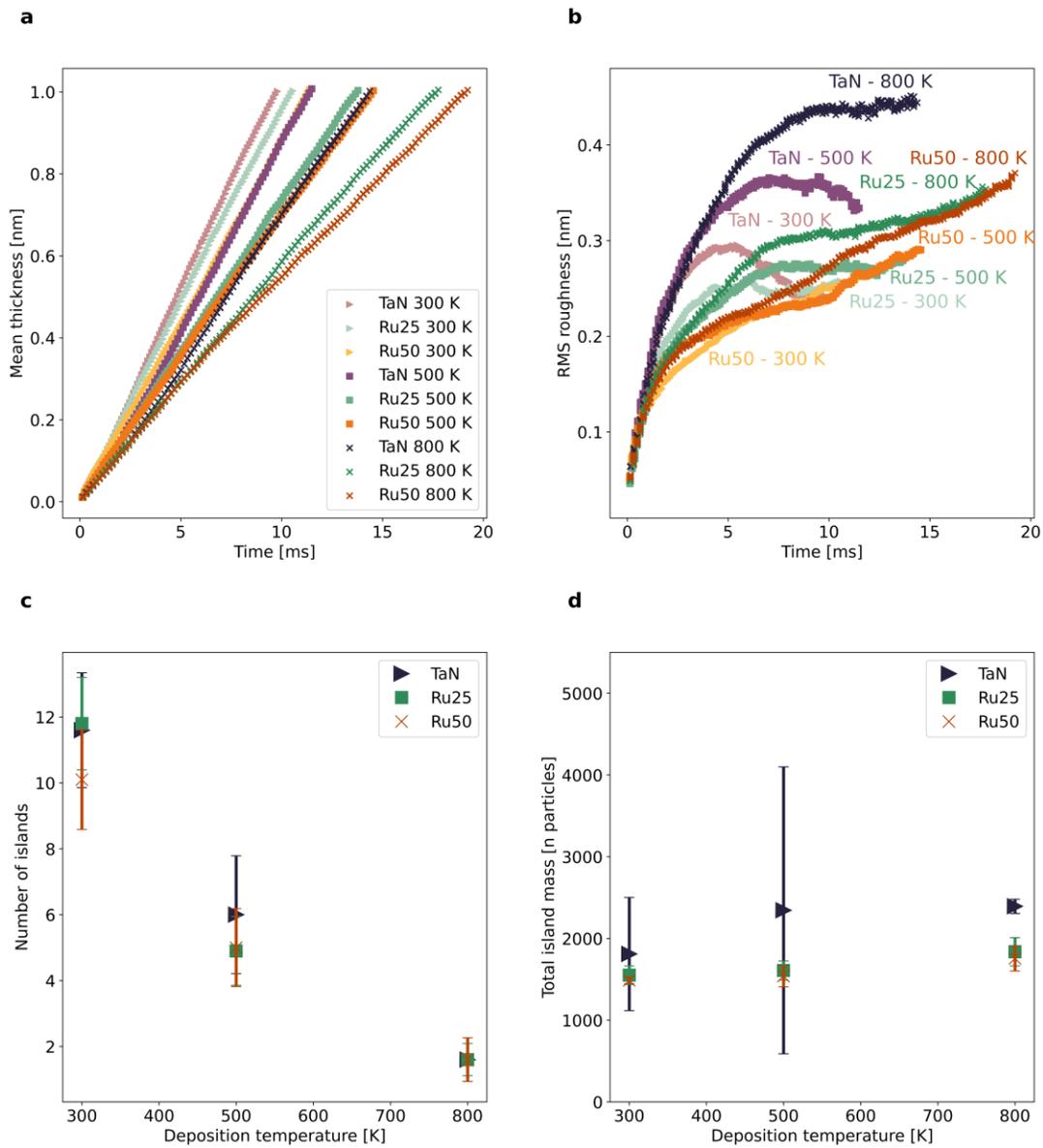

**Figure 2.** Time evolution of thickness (a), RMS roughness (b), averaged number of islands (c) and averaged total island mass (d) in deposition processes at different temperatures (300 K, 500 K and 800 K) at P = 0.5 Pa. Subfigures a) and b) present data from 9 independent simulations each. Subfigures c) and d) show the averaged results from 10 simulations per data point, with error bars as the standard deviation, totaling 90 independent simulations in each figure. Different

categories of colors represent different substrates (purple for TaN, blue for Ru25 and brown for Ru50), with darker colors for higher temperatures.

Figure 2a shows the temporal evolution of the mean thickness. The same categories of colors are chosen for each material (purple for TaN, blue for Ru25 and brown for Ru50), with darker colors for higher temperatures. The growth rate is significantly influenced by partial pressure, which increases the flux of atoms impacting the surface. Specifically, raising the pressure from 0.1 Pa (Figure S9a) to 100 Pa (Figure S10c) reduces the time required to grow a 1 nm thick layer from 100 ms to 0.1 ms. Conversely, higher temperatures retard the deposition process. This behavior aligns with the expression for the transition rate of the deposition process ($k_{ads}$), derived from the kinetic gas theory and presented in the Methodology section. The expression demonstrates that an increase in partial pressure raises the transition rate, while higher temperatures reduce it. The results in Figure 2a also demonstrate that substrates with weaker interactions promote a more rapid increase in mean thickness, suggesting a preference for rapid vertical island growth over horizontal growth. Among the substrates studied Ru50 exhibits the smallest increase in film thickness, revealing that stronger Cu-substrate interactions prevent vertical growth and instead promote horizontal growth, which naturally yields a reduced film thickness for the same deposition time.

The temporal evolution of the RMS roughness provides insights into the changes in the surface morphology of the deposited film. This analysis helps to understand the influence of the substrate interaction and growth temperature in the formation of smoother or rougher films. Generally, Cu deposited on TaN exhibits the highest RMS values, Ru25 shows intermediate values, and Ru50 presents the lowest RMS values, as shown in Figure 2b, which is consistent with the different morphologies promoted by each substrate. For example, a Cu film deposited on a TaN substrate at 500K and 800 K has a RMS roughness of 0.33 nm and 0.44 nm; whereas on a Ru50 substrate at the same temperatures, the RMS roughness is 12% (0.29 nm) and 16% lower (0.37 nm).

The substrate influence on the RMS roughness is further illustrated in Figure S11 for various partial pressures with each data point being the average of 10 independent simulations. Figure S11 also shows that elevated deposition temperatures can result in increased RMS values. This increase may be attributed to the coalescence of islands and the deepening of valleys. Specifically, Figure S11b shows that the film deposited on a TaN substrate at 300 K has a mean RMS roughness of 0.24 nm, while the same deposition at 800 K yields an RMS roughness of 0.44 nm. This arises from the weaker interaction of Cu and TaN, which leads to temperature promoting island formation instead of horizontal growth[36].

Tracking the number of islands and their total mass provides information about the growth mode and the film quality. In 2D growth, island formation is limited by the higher energy barrier for upward migration, resulting in a smaller total island mass compared with 3D growth. Consequently, the total island mass reflects Cu accumulation and surface roughness, serving as an indicator of film smoothness. Figure 2c shows the distribution of the number of islands identified after finishing 10 deposition simulations at each temperature. The results show a decrease in the average number of islands with increasing temperature, which is attributed to the merging of islands. While the temperature exerts a strong effect, the substrate employed plays an equally important role. Ru-modified substrates such as Ru25 and Ru50, which have stronger interactions compared to the unmodified case (TaN), generally exhibit fewer islands. Figure S7 illustrates the surface layers and Figure S8 the corresponding height color maps for increasing growth temperatures for the three substrates, revealing a decrease in the number of islands and an increase in their size with higher temperatures. This trend is further supported by Figure S12, which demonstrates that the mean island size grows with rising temperatures across a range of partial pressures (from 0.1 to 100 Pa). Figure 2d presents the averaged total island mass from 10 deposition simulations conducted on TaN, Ru25 and Ru50 substrates at temperatures of 300 K, 500 K and 800 K. This figure quantifies the mass that migrates beyond the mean layer thickness (1 nm). Although no clear correlation with partial pressure is observed (see Figure S9j-S9l and Figure S10j-S10l), the choice of substrate has a clear impact. TaN exhibits a higher total island mass compared to Ru25, while Ru50 shows the lowest values. Temperature influences Cu migration beyond the mean thickness, with TaN showing the most significant effect. Therefore, Ru incorporation improves the quality of the deposited film by hindering the upward migration of atoms, thereby reducing island formation and promoting the morphology needed for conducting Cu films.

*LAYER OCCUPATION AS A MEASURE OF CU MORPHOLOGY*

The flat surface area and layer occupation analysis of deposited films offers additional information regarding the film roughness. While roughness values may be low if both islands and valleys are small, this does not guarantee the presence of large, smooth regions. Measuring the maximum flat surface area provides further quantitative analysis of a deposited film's uniformity. Additionally, the occupation rate per layer reveals how Cu is distributed across the layers, allowing for comparison with an ideally deposited film, where each layer would be fully occupied. Figure 3 shows the evolution of the maximum flat surface area observed across all layers and the occupation rate per layer for Cu films deposited at P = 0.5 Pa on various substrates (TaN, Ru25 and Ru50) at different temperatures (300 K, 500 K and 800 K). The flat surface area for each layer is determined by calculating the difference between the number of occupied sites in consecutive layers $\Delta n = (n_i - n_{i+1})$, then multiplying this difference by the area per site (0.055 nm$^2$).

Figures 3a, 3b, and 3c display how the maximum flat surface area on the surface of the layers evolves over time. Initially, the maximum flat surface areas are nearly 100 nm$^2$, corresponding to the simulation domain, as the surface coverage is minimal with only a few scattered Cu atoms. As islands and layers grow during deposition, the flat surface area gradually decreases as these structures expand, eventually covering the underlying layer. A more rapid reduction in flat surface area is observed for Ru25 and Ru50 compared to TaN. The observed oscillations correspond to the extension of layers (increasing values) and the coverage of underlying layers (decreasing values). The weakly interacting substrate TaN exhibits a slower reduction in flat surface area because Cu atoms tend to form islands rather than completing layers before the deposition of the next layer starts. Higher levels of Ru concentration promote horizontal growth over vertical growth, resulting in more efficient substrate coverage and desired 2D morphology, even at higher temperatures.

Figure S13a, S13b, and S13c show the substrate exposure at different temperatures, defined as the area of the substrate not covered by Cu. Averaged over 10 independent simulations, the data for partial pressures of 0.1 (Figure S13a), 0.5 (Figure S13b) and 1 Pa (Figure S13c) demonstrate that at 800 K, the substrate exposure for TaN (23.76 nm²) is an order of magnitude higher compared to Ru50 (2.48 nm²). At 500 K, TaN shows a substrate exposure of 16.46 nm², while Ru50 achieves complete coverage. Ru50 shows the lowest substrate exposure values, consistent with the promotion of horizontal growth and hence 2D morphology. Figures S13d, S13e, and S13f show that elevated temperatures result in a reduction in the total flat surface area, which represents the sum of the flat surface areas across all layers. This reduction is more pronounced for the TaN

substrate compared to Ru25, and is most significant when compared to Ru50. The decrease in total flat surface area occurs due to island formation, which contributes to covering the flat surfaces. This explains why Ru50 exhibits the highest flat surface area.

Figures 3d, 3e, and 3f display the occupation rate per layer at the end of the simulation at 300 K, 500 K, and 800 K, respectively. The data shows that for TaN, the occupation rate of the first layer decreases significantly from 94% at 300 K to 77% at 800 K, suggesting that island formation becomes more favorable at higher temperatures. In contrast, the impact of temperature on Ru25 and Ru50 is less pronounced. Specifically, occupation rate for Ru25 decreases from 99% to 93%, while for Ru50 it drops from 100% to 97%, showing that these substrates hinder island formation independently from the deposition temperature. These findings are consistent with the observations presented in Figure S13a-S13c regarding substrate exposure.

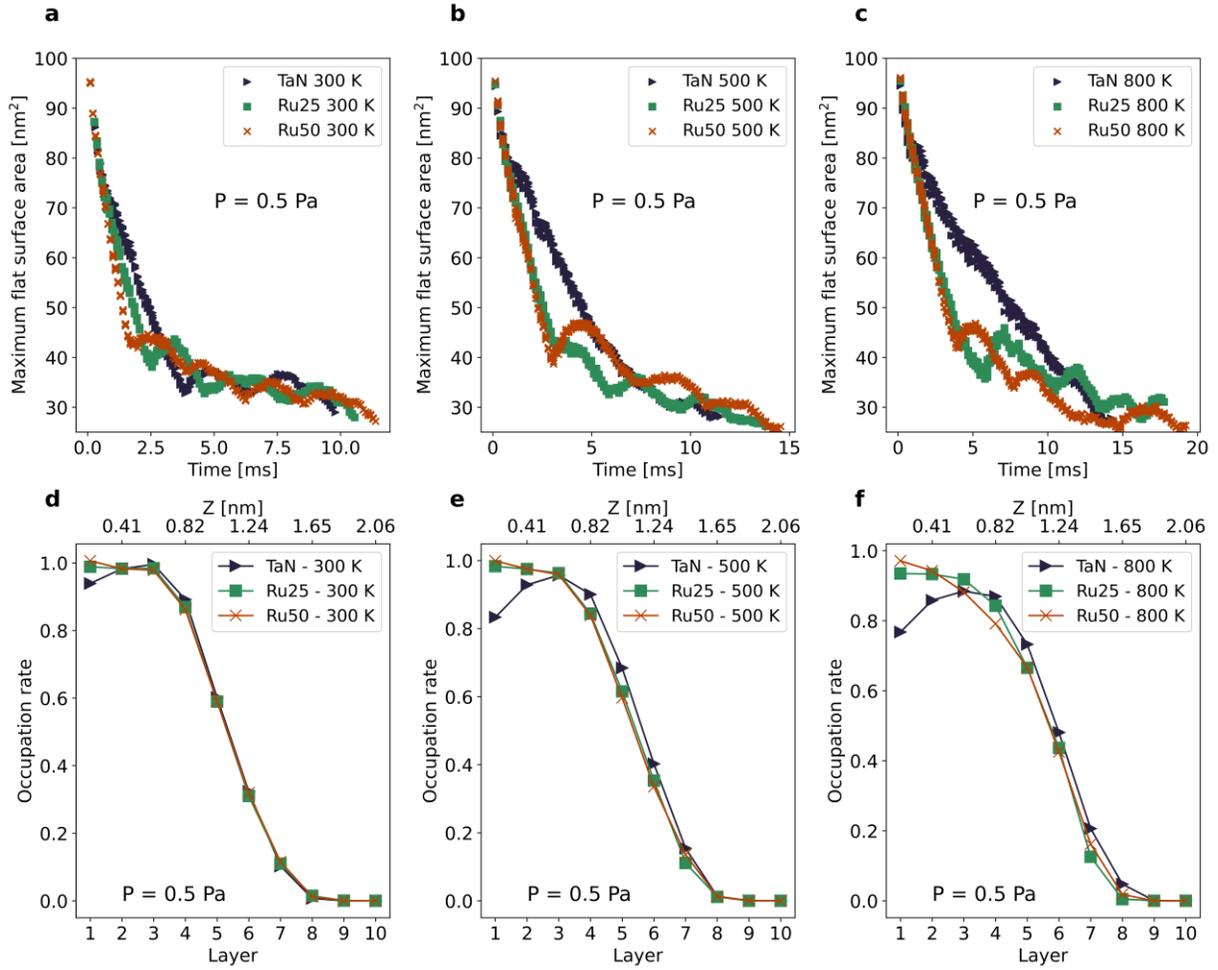

**Figure 3.** Time evolution of the maximum flat surface area across all layers for Cu deposition simulations on various substrates (TaN, Ru25 and Ru50) at different temperatures: 300 K (a), 500 K (b) and 800 K (c). The occupation rate per layer at the conclusion of the simulation for the same substrates at temperatures: 300 K (d), 500 K (e) and 800 K (f). All simulations were conducted at a partial pressure of P = 0.5 Pa. The lines between points are included to guide the eyes.

*DEFECT DENSITY IN DEPOSITED CU*

Analyzing the frequency of Cu atoms with a specific number of neighbors allows us to estimate the defect density within the film. In perfect bulk Cu, as shown in Figure 1a, each Cu atom has twelve nearest neighbors: six in the same plane, three in the layer above and three in the layer below. Atoms in the top and bottom layers, however, can have a maximum of nine nearest neighbors (excluding the substrate). Thus, in an ideal crystal, all atoms would have either nine or twelve neighbors. Deviations from these coordination numbers indicate the presence of defects such as vacancies. In our case, if the first and last layers were fully occupied, we would observe 3,656 Cu atoms with nine nearest neighbors. Figure 4 presents the frequency distribution of Cu atoms based on their number of nearest neighbors, calculated after the mean layer thickness reached 1 nm for the three substrates. At all three temperatures—300 K (Figure 4a), 500 K (Figure 4b), and 800 K (Figure 4c)—the number of Cu atoms with 9 neighbors is lower when Cu is deposited on TaN. In contrast, the number of atoms with fewer than 9 or 10 and 11 nearest neighbors is higher compared to the other substrates. Conversely, the Ru50 shows the highest frequency of 9 nearest neighbors and lowest frequency of cases deviating from this, that is atoms with 8, 10 or 11 nearest neighbors, while Ru25 is an intermediate case. These results suggest that Ru50 exhibits the lowest number of defects, as further evidenced in Figure 4d. This figure shows the frequency of Cu atoms with 9 nearest neighbors, averaged over 10 simulations, across a range of temperatures. Additionally, the data indicate that increasing temperature raises the number of Cu atoms with 9 nearest neighbors (Figure 4d) and 12 nearest neighbors (Figure 4a-4c), signifying a more compact layer structure.

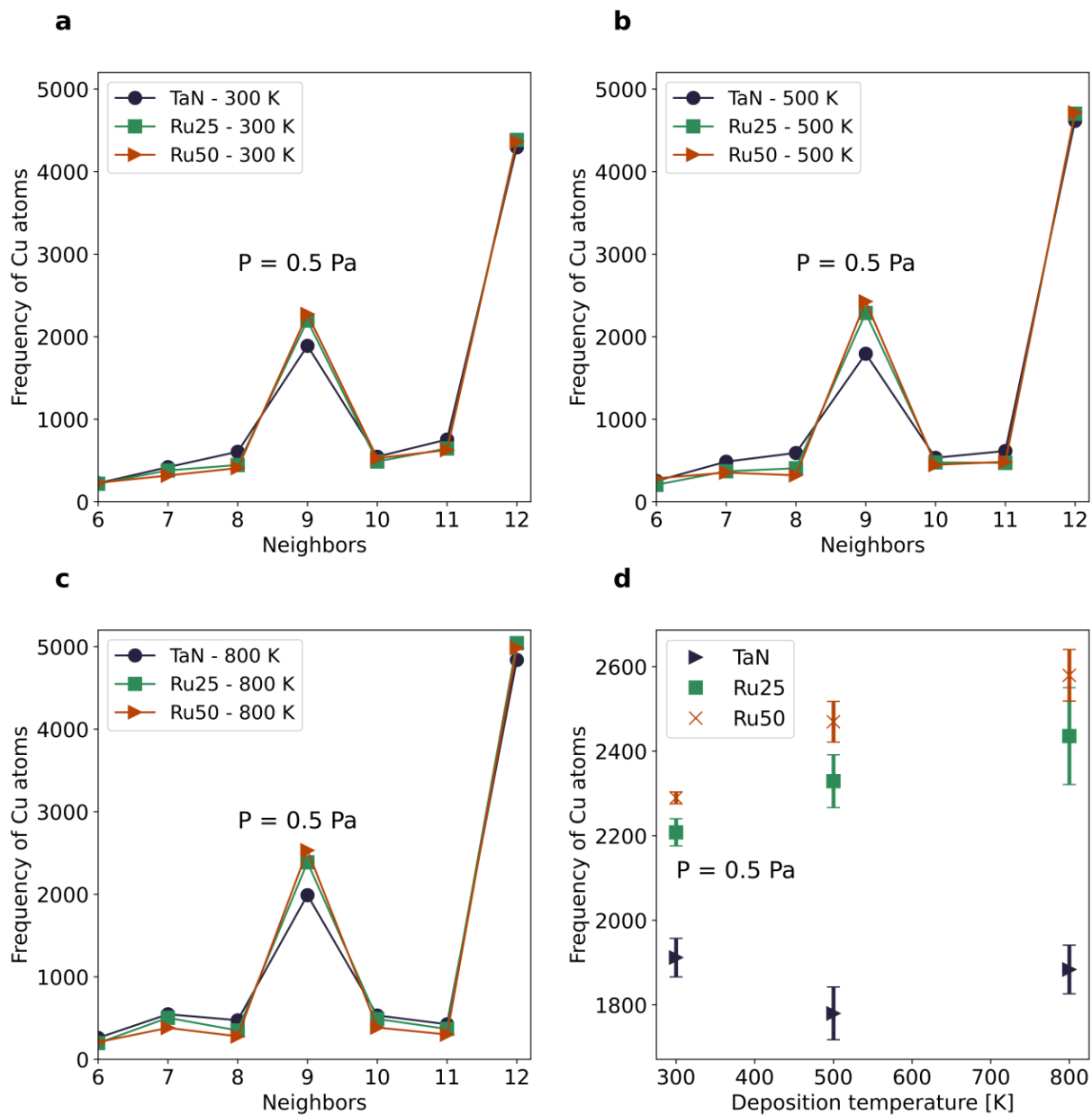

**Figure 4.** Frequency distribution of Cu atoms by the number of nearest neighbors for TaN, Ru25 and Ru50 substrates at a) 300 K, b) 500 K and c) 800 K. d) Frequency of Cu atoms with 9 nearest neighbors, averaged over 10 simulations, for TaN, Ru25 and Ru50 substrates across temperatures of 300 K, 500 K and 800 K. All simulations were conducted at a partial pressure of P = 0.5 Pa.

*EFFECT OF VACUUM ANNEALING ON FILM MORPHOLOGY AND QUALITY*

Thermal vacuum annealing is a damage-healing process used in thin film deposition to reduce defects and recrystallise from an amorphous film to a crystalline film, thereby optimizing the electrical and mechanical properties of the material for high-performance applications[65]. We explore in this section the role of vacuum annealing on the morphology and quality of as-deposited Cu films. The temperature of annealing facilitates the migration of undercoordinated Cu atoms to more stable positions, increasing their CN. This include both upward migration, when not hindered by substrate interaction, and downward migration from the uppermost layers that can promote horizontal growth. For weakly interacting TaN substrates, this leads to the deepening of valleys and an increase in CN for atoms within islands, as shown in Figure 5a and 5b. In contrast, for strongly interacting substrates, such as Ru50, surface flattening is expected, since the high activation barriers means that Cu atoms migrating downward rarely return to upper layers (see Figure 5c and 5d). To investigate the impact on crystal quality and the potential for enhancing layer uniformity and promoting a denser atomic arrangement, Cu deposited on TaN, Ru25 and Ru50 (the same simulations discussed above) at 500 K and P = 0.5 Pa are annealed in vacuum for 2.5 million kMC steps at three temperatures: 300 K, 500K and 800 K. We examined changes in occupation rate (Figure 6a, 6b and 6c) and the frequency distribution of Cu atoms with a specific number of nearest neighbors (Figure 6d, 6e and 6f).

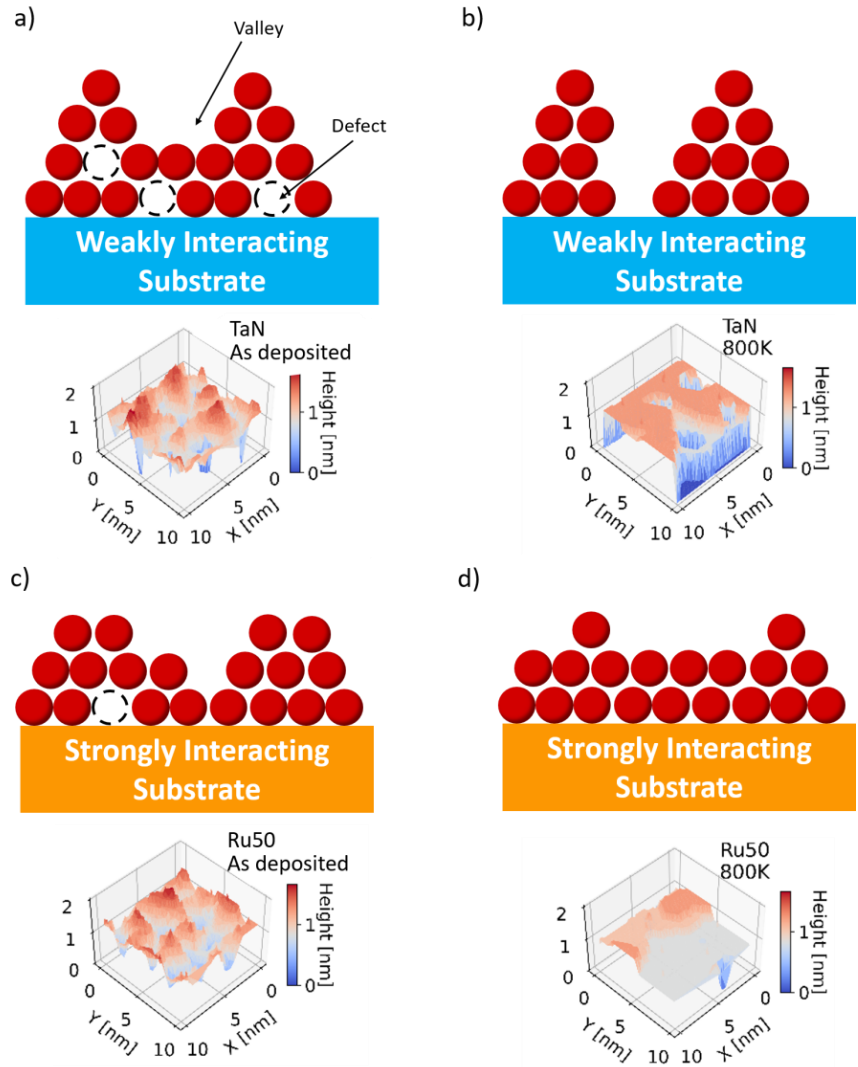

**Figure 5.** Schematic and simulation of a Cu film deposited on a) a weakly interacting substrate (TaN) before annealing, and b) after annealing. Schematic and simulation of a Cu film deposited on c) a strongly interacting substrate (Ru50) before annealing, and d) after annealing. The 3D images depict the surface of the deposited Cu film, with blue indicating low regions (valleys, pits) and red representing elevated regions (islands).

Ru incorporation significantly affects the rearrangement of Cu atoms within the film at different temperatures, as illustrated in Figures 6a (TaN), 6b (Ru25) and 6c (Ru50). Annealing at 300 K shows minimal change compared to the as-deposited films, while annealing at 500 K has a minor impact. For TaN, the occupation rate of the first layer decreases from 84% after deposition to 68% after annealing at 800 K. This indicates that annealing as-deposited Cu on TaN drives upwards Cu migration and leads to deepening of valleys (that is the space between islands is enhanced - see the surface in Figure S14d and the schematic in Figure 5a and 5b) which increases the island uniformity. In contrast, Ru25 shows significant improvement (see Figure S14h), with the occupation rate of the first layer hardly decreasing, from 99% to 93%, after annealing, as shown in Figure 6b. In this case, Figure 6b shows how some atoms from layers below 5 also migrate upward. Conversely, for TaN, the occupation rates for the layers five and six increase due to downward migration from layers above 6 and upward migration from near the substrate (Figure 6a). Ru50 performs best, as can be seen in Figure S14l, maintaining a 100% occupation rate in the first layer and increased occupation rates up to the fourth layer (see Figure 6c). This highlights the ability of Ru incorporation to minimize substrate exposure, even at high annealing temperatures, as Figure 7b illustrates. Notably, for Ru25 and Ru50, many atoms from layers above layer five migrate downwards, reducing both the occupation rate of layers above the sixth layer and the extent of islanding. This downward migration suggests that annealing facilitates Cu redistribution, filling layers closer to the substrate, promoting 2D growth and denser film formation, and thus enhancing crystal quality.

Cu films deposited and then annealed on Ru-TaN substrates show a reduced number of undercoordinated Cu atoms or those with no nearest neighbors compared to TaN substrate, while increasing the frequency of atoms with 9 neighbors, particularly at higher annealing temperatures, as shown in Figures 6d (TaN), 6e (Ru25) and 6f (Ru50). This change suggests a denser atomic arrangement, notably for the case of Ru50.

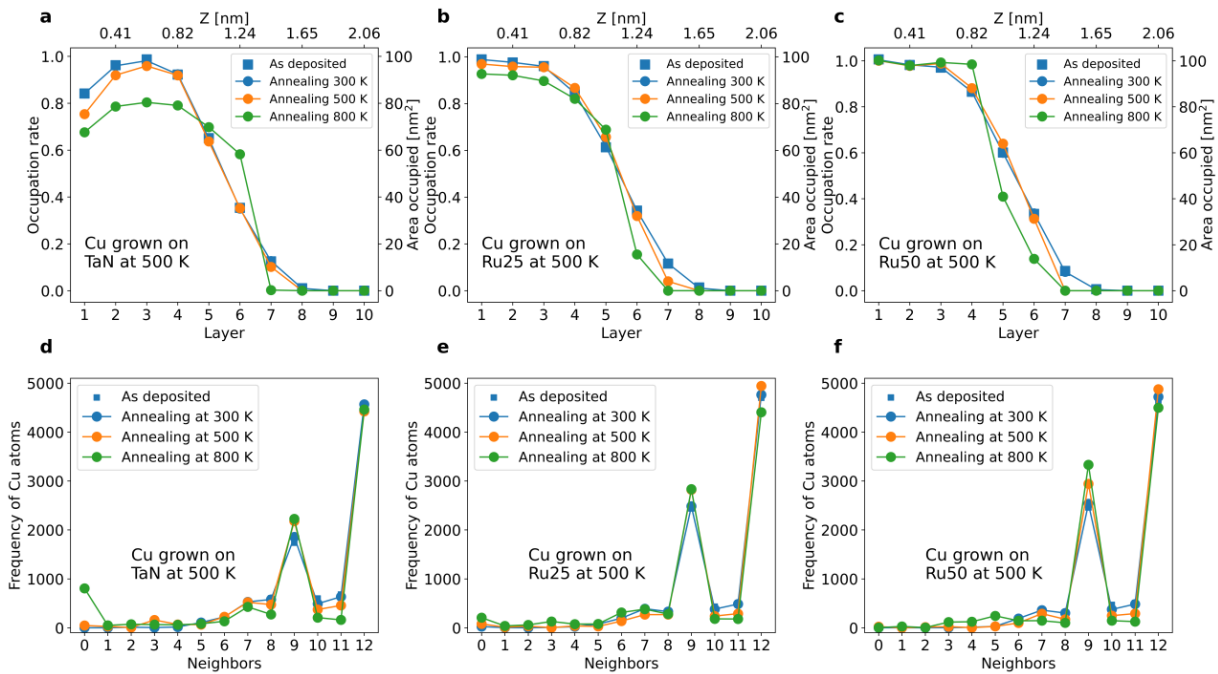

**Figure 6.** The occupation rate per layer at 300 K, 500 K and 800 K annealing process for TaN (a), Ru25 (b) and Ru50 (c). Frequency distribution of Cu atoms by the number of nearest neighbors at 300 K, 500 K and 800 K vacuum annealing for TaN (d), Ru25 (e) and Ru50 (f).

The influence of Ru incorporation in the substrate on the formation of the maximum flat surfaces during annealing is shown in Figure 7d. While increased annealing temperature clearly promotes larger flat surface areas, the effect of the substrate is less distinct, as all substrates achieve similar surface areas. However, when examining the post-annealing surfaces of Cu deposited at 500 K on TaN, Ru25, and Ru50 (Figure S14), it becomes clear that Ru incorporation limits valley formation and substrate exposure (see Figure 7b). On the TaN substrate, annealing flattens the surface, but also deepens valleys, leading to increased substrate exposure and more pronounced islands (see Figure S14a–S14d). A similar effect is observed for Ru25 (Figures S14e–S14h), where surface flattening is accompanied by reduced substrate exposure due to stronger substrate interactions. Annealing as-deposited Cu on Ru50 (Figures S14i–S14l), which has the strongest substrate interaction, results in the lowest substrate exposure and the smoothest surfaces, with fewer valleys. Figure S15 provides the corresponding height color maps for the layer surfaces shown in Figure S14.

Analyzing RMS roughness further supports the conclusion that Ru incorporation in combination with a vacuum anneal results in horizontal growth, primarily by limiting valley formation. Figure 7f shows the RMS roughness of Cu deposited at 500 K on different substrates across the range of annealing temperatures. For TaN, annealing causes a significant increase in RMS roughness, from 0.28 nm at 300 K to 0.49 nm at 800 K. In contrast, Ru25 shows a smaller increase in RMS roughness, from 0.28 nm at 300 K to 0.32 nm at 800 K. Remarkably, for Ru50, the RMS roughness decreases with temperature, from 0.25 nm at 300 K to 0.17 nm at 800 K. Comparing the effects across substrates for Cu deposited at 300 K and 500 K (Figure S16) further highlights how increased Ru content improves RMS roughness, which is highly relevant for BEOL processing of Cu layers.

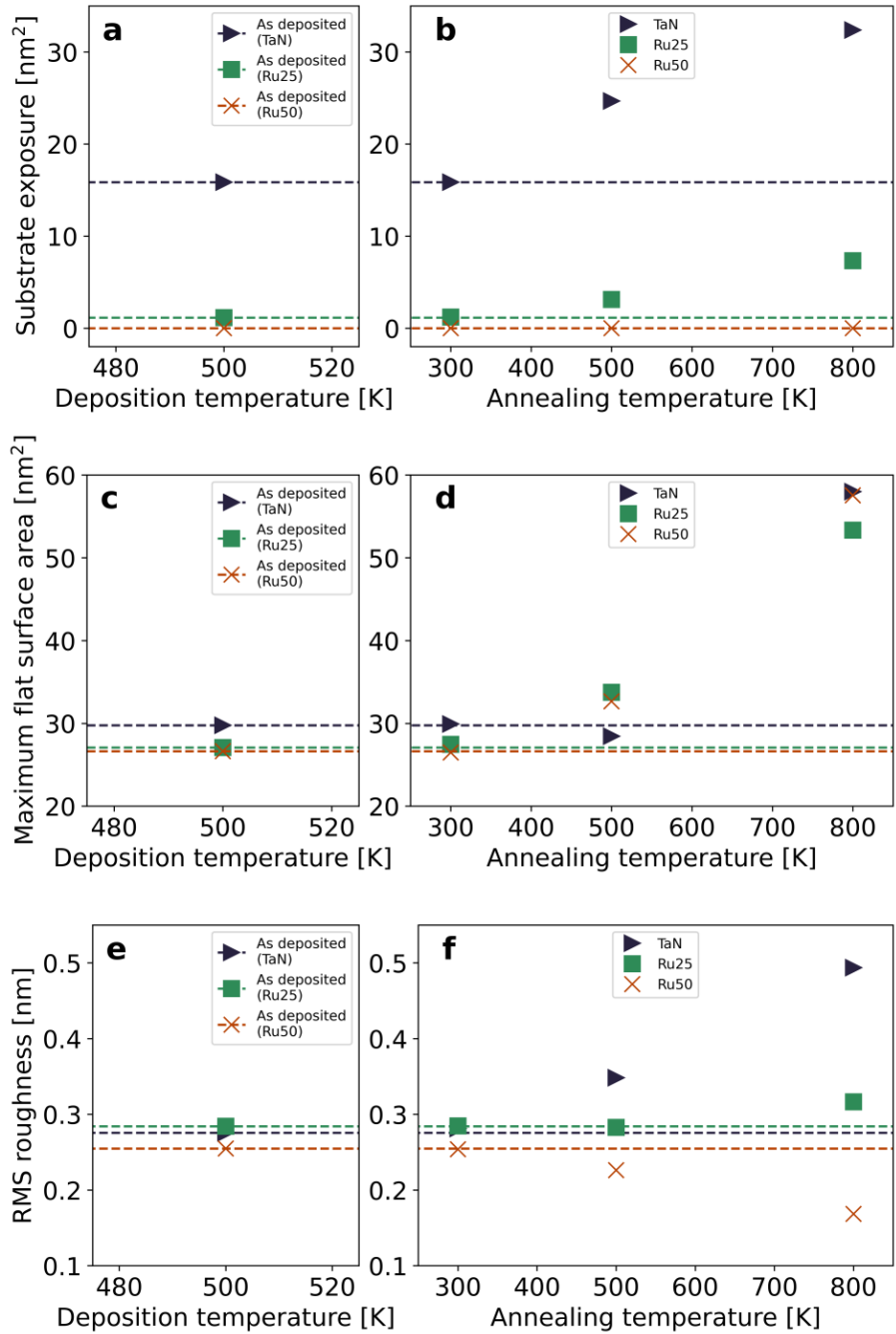

**Figure 7:** Substrate exposure for as Cu deposited at 500 K (a) and after annealing (b). Maximum flat surface area for as Cu deposited at 500 K (c) and after annealing (d). RMS roughness for as Cu deposited at 500 K (e) and after annealing (f). Cu is deposited at 500 K and P = 0.5 Pa on TaN, Ru25 and Ru50 and then annealed at 300 K, 500 K and 800 K. Horizontal dashed lines indicate reference values for the as-deposited films for each substrate.

CONCLUSION

Finding combined barrier/liner materials that free up volume and promote deposition of smooth Cu interconnects while hindering island formation is key for the miniaturization of electronic devices. We have implemented a kinetic Monte Carlo simulator that uses DFT data to model Cu deposition and annealing and predict how the deposition temperature, pressure and annealing determine the morphology of deposited Cu films on TaN and Ru-modified TaN substrates. Our results demonstrate that Ru-modified TaN is a strong candidate for a combined Cu barrier and liner layer to promote improved Cu film deposition in terms of film quality and uniformity, which is crucial for the miniaturization of electronic devices to sub-nm. Our kMC Cu deposition simulations on TaN substrates allow us to predict the effects of Ru incorporation in the substrate, temperature and pressure on the growth and morphology of deposited Cu. We furthermore study the effect of vacuum annealing on Cu morphology. This enables a new paradigm for predictive atomistic simulations of metal deposition processes on technologically important substrates, with particular target morphologies, which can be used to select suitable barrier/liner layers for controlled deposition of interconnect metal films and it can be readily extended to, e.g. catalysis to predict the morphology of deposited metal nanoparticles for catalytically important reactions, thus signposting the optimal nanoparticle deposition conditions (substrate, temperature, pressure, annealing).

Our results demonstrate that incorporating Ru into TaN substrates significantly improves their performance in promoting the deposition of conducting Cu by favoring a 2D morphology with horizontal film growth over a 3D non-conducting island morphology. At the highest level of Ru incorporation (50%) and a deposition temperature of 800 K the surface roughness is reduced by over 16%, substrate exposure decreases by an order of magnitude and compactness is enhanced by reducing the number of undercoordinated Cu atoms. The resulting Cu layers exhibit superior quality, even at elevated temperatures, which typically increase RMS roughness and substrate exposure during film deposition. These findings suggest that the increased substrate interaction strength from Ru incorporation promotes horizontal (2D) over vertical (3D) growth, which is crucial for achieving high-quality Cu interconnects. Vacuum annealing studies (from 300 K to 800 K) reveal that annealing at higher temperatures can further tune the morphology post deposition, depending on the substrate employed. While elevated annealing temperatures can be advantageous in reducing vacancy defects, decreasing the number of uncoordinated atoms, and promoting flatter surfaces on Ru-TaN, they can further promote Cu island formation on TaN by promoting upwards Cu migration that deepens valleys and increases substrate exposure, which in turn raises RMS roughness. This is expected given the weaker interaction between Cu and TaN as compared to Ru-TaN, which alters the growth mode to promote island formation. However, Ru incorporation mitigates these negative effects by increasing the Cu substrate interaction and, at the highest level of incorporation (50%), can even produce beneficial effects by not only minimizing the number of uncoordinated Cu atoms but also reducing the RMS roughness. This demonstrates that Ru modification can also contribute to the stability of the layers under annealing conditions.

ASSOCIATED CONTENT

**Supporting Information**.

The following files are available free of charge.

Supporting Information showing: estimation of activation energies estimation, 3D layer surfaces and color maps, partial pressure comparison, RMS roughness, island sizes, substrate exposure and annealing impact on layer surfaces and roughness (PDF)

AUTHOR INFORMATION

**Corresponding Author**

Samuel Aldana - Tyndall National Institute, University College Cork, Lee Maltings, Dyke Parade, Cork T12 R5CP, Ireland. https://orcid.org/0000-0002-1719-605X

E-mail: samuel.delgado@tyndall.ie

Michael Nolan - Tyndall National Institute, University College Cork, Lee Maltings, Dyke Parade, Cork T12 R5CP, Ireland. https://orcid.org/0000-0002-5224-8580

E-mail: michael.nolan@tyndall.ie

**Author Contributions**

S. A. developed and conducted the kMC simulations. C-L N. conducted DFT and MD simulations. S.A., C-L N. and M. N. analyzed the data. M. N. supervised the project. The manuscript was written through contributions of all authors. All authors have given approval to the final version of the manuscript.

**Funding Sources**

The ASCENT+ Access to European Infrastructure supported MN and SD for Nanoelectronics Program, funded through the EU Horizon Europe programme, grant no 871130. CLN and MN were supported through the Science Foundation Ireland SFI-NSF China Partnership program, grant number NSF/2018/e2222.

**Table of Contents graphic**

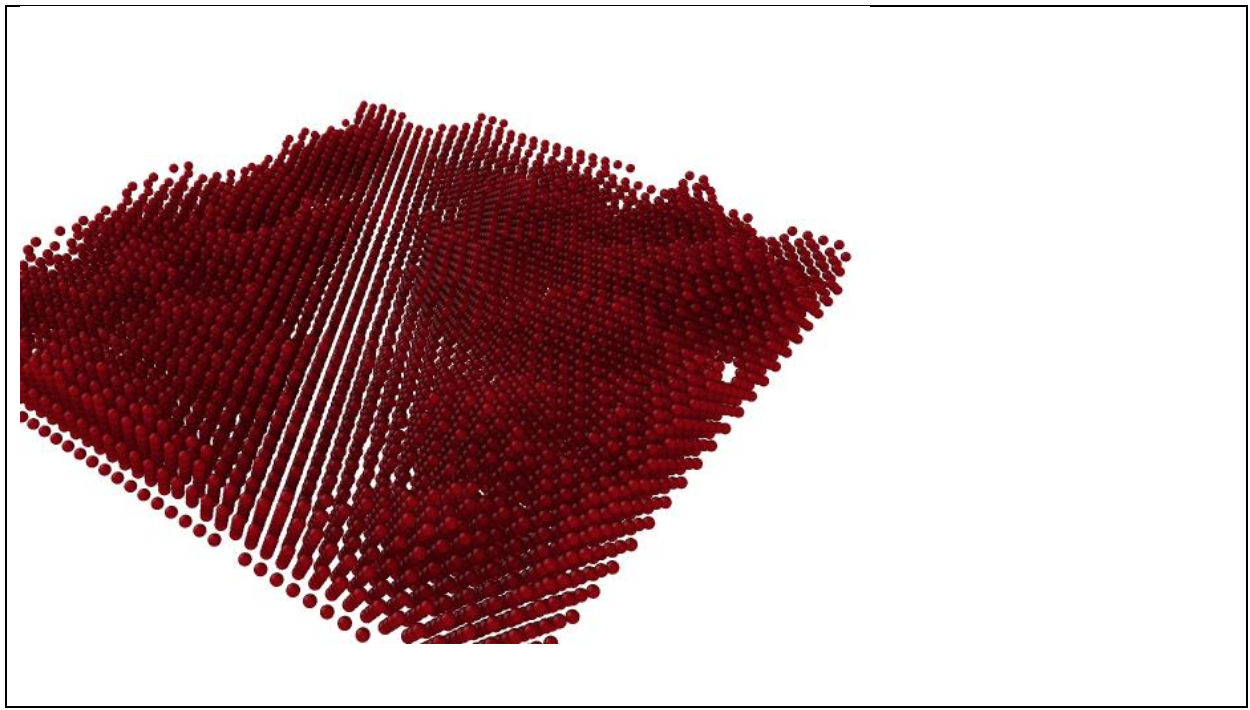